\documentclass[runningheads]{llncs}
\usepackage{graphicx}
\begin{document}
\title{A Deep Regression Model for Seed Localization in Prostate Brachytherapy}
%
%
\author{Yading Yuan \and
Ren-Dih Sheu \and
Luke Fu \and
Yeh-Chi Lo}
\authorrunning{Y. Yuan et al.}
%
\institute{Department of Radiation Oncology\\
Icahn School of Medicine at Mount Sinai\\
New York NY 10029, USA\\
\email{yading.yuan@mssm.edu}}
\maketitle              
\begin{abstract}
Post-implant dosimetry (PID) is an essential step of prostate brachytherapy
that utilizes CT to image the prostate and allow the location and
dose distribution of the radioactive seeds to be directly related
to the actual prostate. However, it it a very challenging task to
identify these seeds in CT images due to the severe metal artifacts
and high-overlapped appearance when multiple seeds clustered together.
In this paper, we propose an automatic and efficient algorithm based
on 3D deep fully convolutional network for identifying implanted seeds
in CT images. Our method models the seed localization task as a supervised
regression problem that projects the input CT image to a map where
each element represents the probability that the corresponding input
voxel belongs to a seed. This deep regression model significantly
suppresses image artifacts and makes the post-processing much easier
and more controllable. The proposed method is validated on a large
clinical database with 7820 seeds in 100 patients, in which 5534 seeds
from 70 patients were used for model training and validation. Our
method correctly detected 2150 of 2286 (94.1\%) seeds in the 30 testing
patients, yielding 16\% improvement as compared to a widely-used commercial
seed finder software (VariSeed, Varian, Palo Alto, CA).

\end{abstract}
\section{Introduction}
With estimated 174,650 new cases and 31,620 deaths in 2019, prostate
cancer remains the most common type of cancer diagnosed in men in
the US~\cite{Siegel_2019}. Seed implant brachytherapy, which involves
permanent implantation of radioactive sources (seeds) within the prostate
gland, is the standard option for low and intermediate risk prostate
cancer~\cite{Chin_2017}. Despite various improvements in planning
and seed delivery, the actual radiation dose distribution may deviate
from the plan due to various factors such as needle positioning variations,
prostate deformation, seed delivery variations and seed migration.
Therefore, post-implant dosimetry (PID) is recommended to assure the
quality of the implantation and to establish the relationship between
radiation dose and clinical outcomes~\cite{Stock_2002}. PID is typically
performed at day 30 following implantation that utilizes CT to image
the implanted area, from which prostate and surrounding organs at
risk (OARs) are outlined and seed locations are identified.

\begin{figure}[t!]
\includegraphics[width=1\textwidth]{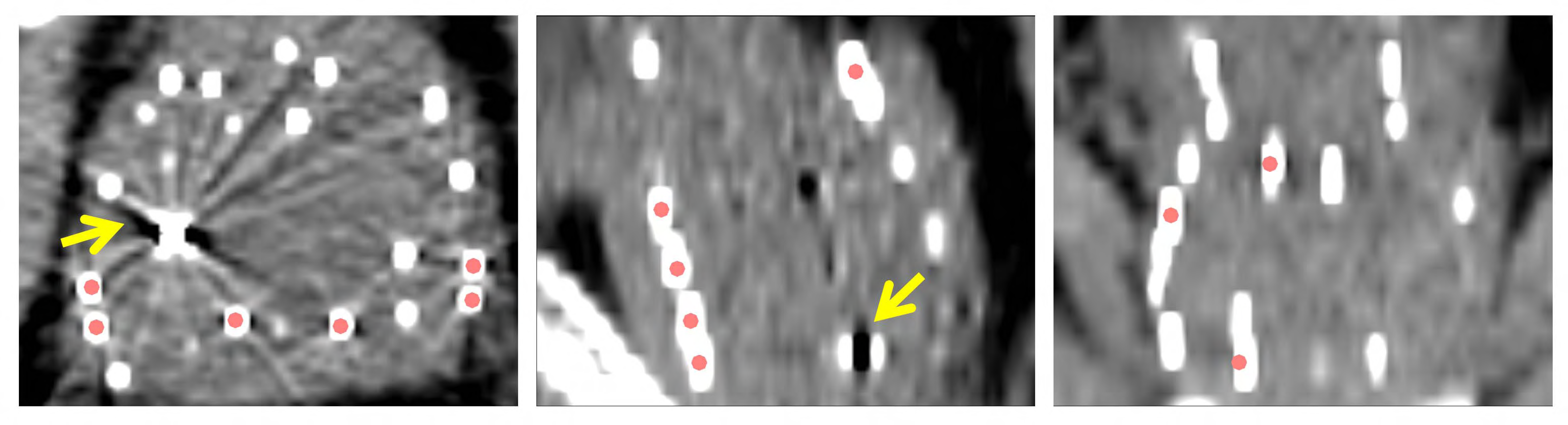} \caption{An example of seed appearance in CT images in axial (left), sagittal
(middle) and coronal (right) view, respectively. Yellow arrows indicate
the metal artifacts and red dots represent mannually annotated seed
locations. Clustered seeds can be clearly seen in saggital and coronal
views, as indicated by the blue arrows.}
\label{fig-challenges}
\end{figure}

Accurate localization of implanted seeds is essential to quantify
the dose distribution to those organs. However, manual identification
of these seeds is time consuming given a large number of seeds implanted,
typically taking 10 - 20 minutes to identify 60 to 100 seeds per patient.
Therefore, accurate and automated methods for seed localization are
of great demand. While the radio-opaque seeds appear with high contrast
on the CT images, automatic seed localization is in practice a challenging
task due to the following two unique characteristics, as shown
in Fig. \ref{fig-challenges}. Firstly, the presence of fudicial markers introduces
severe metal artifacts on CT images, which significantly increases
the complexity of seed identification. Secondly, due to seed delivery
variations and seed migration, some implanted seeds are very close
to each other to form seed clusters. This highly-overlapped appearance
make it hard to identify individual seed on CT images.

Several automatic approaches have been developed to localize seeds
in CT images such as geometry-based recognition method~\cite{Liu_2003}
and Hough transform ~\cite{Holupka_2004}. Recently, Nguyen et al.~\cite{Nguyen_2014} proposed a cascaded method that involves thresholding
and connected component analysis as initial detection of seed candidates,
and followed by a modified k-means method to separate groups of seeds
based on a priori intensity and volume information. Zhang et al.~\cite{Zhang_2017}
employed canny edge detection and an improved concave points matching
to separate touching seeds after gray-level-histogram based thresholding.
All these methods use hand-crafted features that require specialized
domain knowledge. Meanwhile, sophisticated pre- and post-processing
steps are usually introduced to facilitate the seed localization procedure.
As a result, the evaluation of these methods was mainly conducted
with physical phantom or small amount of clinical cases.

Recently, deep convolutional neural networks (CNNs) have become popular
in medical image analysis~\cite{Litjens_2017} and have achieved
state-of-the-art performance in various medical image computing tasks
such as lung nodule detection~\cite{Setio_2017}, gland instance
segmentation in histology images~\cite{Xu_2017}, liver and tumor
segmentation~\cite{Bilic_2019}, skin lesion segmentation~\cite{Yuan_2017}
and classification~\cite{Yu_2017}. Due to the capability of learning
hierarchical features directly from raw image data, CNNs usually yield
better generalization performance especially when evaluating on a
large scale of dataset.

Enlightened by the latest advances in deep learning research, we propose
a novel framework based on deep CNNs to automatically localize the
implanted seeds in 3D CT images. Our contributions in this paper are
three fold. Firstly, we model seed localization as a regression problem
and introduce a fully automated solution by leveraging the discriminative
power of deep CNNs. To the best of our knowledge, this is the first
attempt of using deep neural networks to tackle this challenging task.
Secondly, instead of directly predicting the seed coordinates in 3D
space, we design a probability map of seed locations to account for
the uncertainty of manual identification, which improves the robustness
of model prediction. Finally, we evaluated the proposed method on
a large clinical database with $7820$ seeds in 100 patients, and
compared the results with a commercial seed finder software (VariSeed,
Varian, Palo Alto, CA).

\begin{figure}[t!]
\includegraphics[width=1\textwidth]{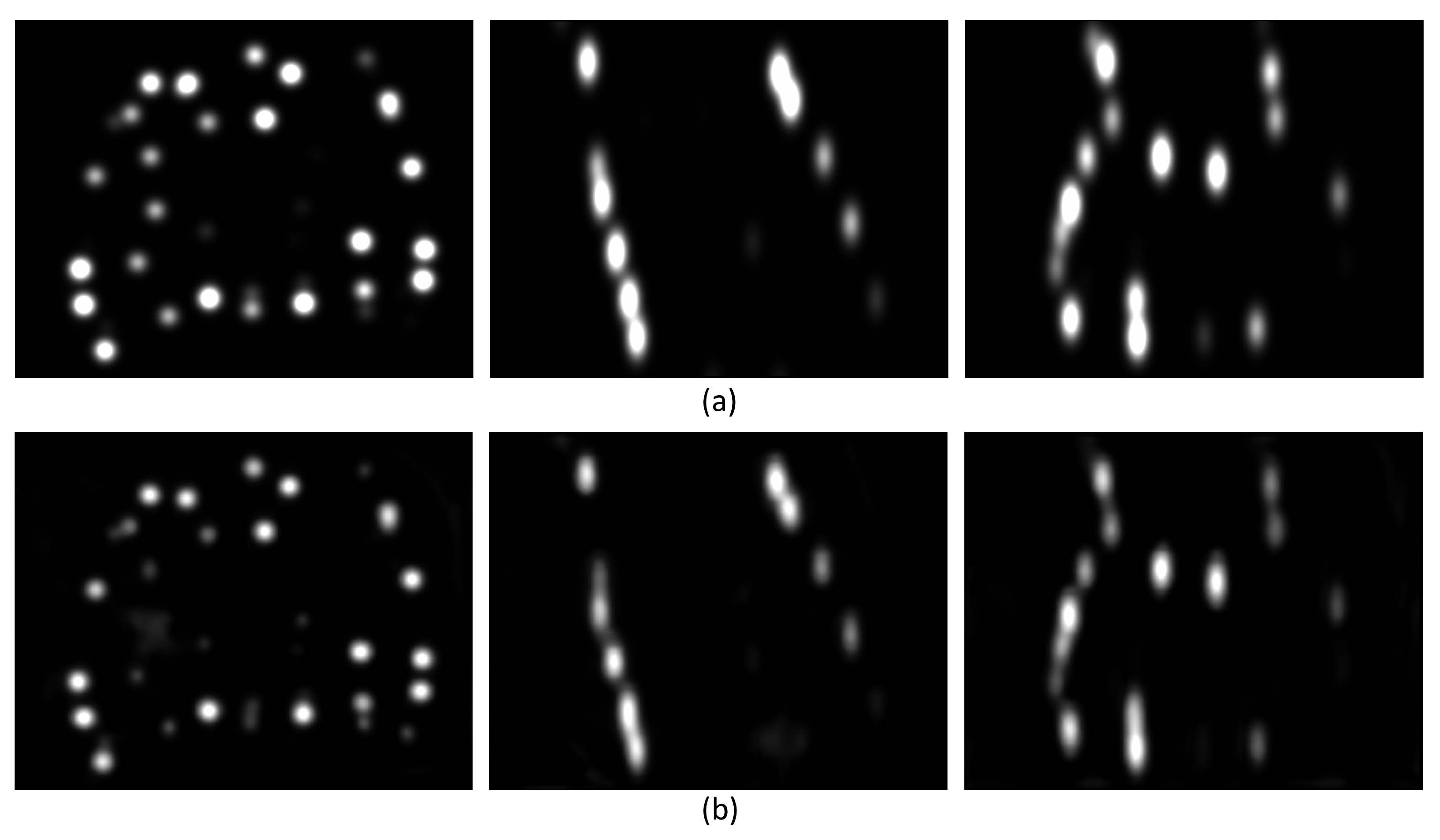} \caption{(a) The target probability maps created from the dot manual annotations
in Figure 1. (b) The corresponding predicted probability maps inferred
from the proposed deep regression network.}
\label{fig-model}
\end{figure}

\section{Methodology}

\subsection{Deep Regression Model}

As shown in Fig. \ref{fig-challenges}, the ground truth is provided
as dot annotations, where each dot corresponds to one seed. However,
considering the seed has a finite dimension (about 0.8 mm in diameter
and 4.5 mm in length), any dot annotation should be considered as correct
as long as it's located on the seed. As a result, a large variation
can be observed in the ground truth in terms of the annotation positions
on the seeds, which makes it unnecessarily challenging and prone to
overfitting if the exact annotation positions are directly used as
learning target. Instead, we convert the discrete dot annotations
into a continuous probability map $P(\vec{x})$ ($\vec{x}\in R^{3}$)
and cast the seed localization task as a supervised regression problem
that learns a mapping between a 3D CT image set $I(\vec{x})$ and
$P(\vec{x})$, denoted as $\hat{P}(\vec{x},\,\vec{w})=F(I(\vec{x}),\,\vec{w})$
for $\hat{P}(\vec{x})$ the inferred probability map and $\vec{w}$
the learned parameters (weights).

For each training image $I_{i}(\vec{x})$ that is annotated with a
set of 3D points $\vec{C}_{i}=\{C_{1},\ldots,C_{N(i)}\}$, where $N(i)$
is the total number of seeds annotated by the user, we define the
ground truth probability map to be a kernel density estimation based
on the provided points:
\begin{equation}
\forall\vec{x}\in I_{i},\;P_{i}(\vec{x})=\sum_{C\in\vec{C}_{i}}\mathcal{N}(\vec{x};\,C,\Sigma),\;\;\;\;\;\Sigma=\left[\begin{array}{ccc}
\sigma_{x}^{2} & 0 & 0\\
0 & \sigma_{y}^{2} & 0\\
0 & 0 & \sigma_{z}^{2}
\end{array}\right].
\end{equation}
Here $\vec{x}$ denotes the coordinates of any voxel in image $I_{i}$,
and $\mathcal{N}(\vec{x};\;C,\Sigma)$ is the normalized 3D Gaussian
kernel evaluated at $\vec{x}$, with the mean at the user annotation
$C$ and a diagonal covariance matrix $\Sigma$. Considering the physical
seed dimension and the magnification effect during CT imaging, we
fixed $\sigma_{x}=\sigma_{y}=1\,mm$ and $\sigma_{z}=2\,mm$ in our
study. Figure \ref{fig-model} (a) shows several examples of the probability
map that are created from the dot manual annotations in Fig. \ref{fig-challenges},
and (b) are the corresponding predicted maps inferred from the proposed
deep regression model.

\begin{figure}[t!]
\includegraphics[width=1\textwidth]{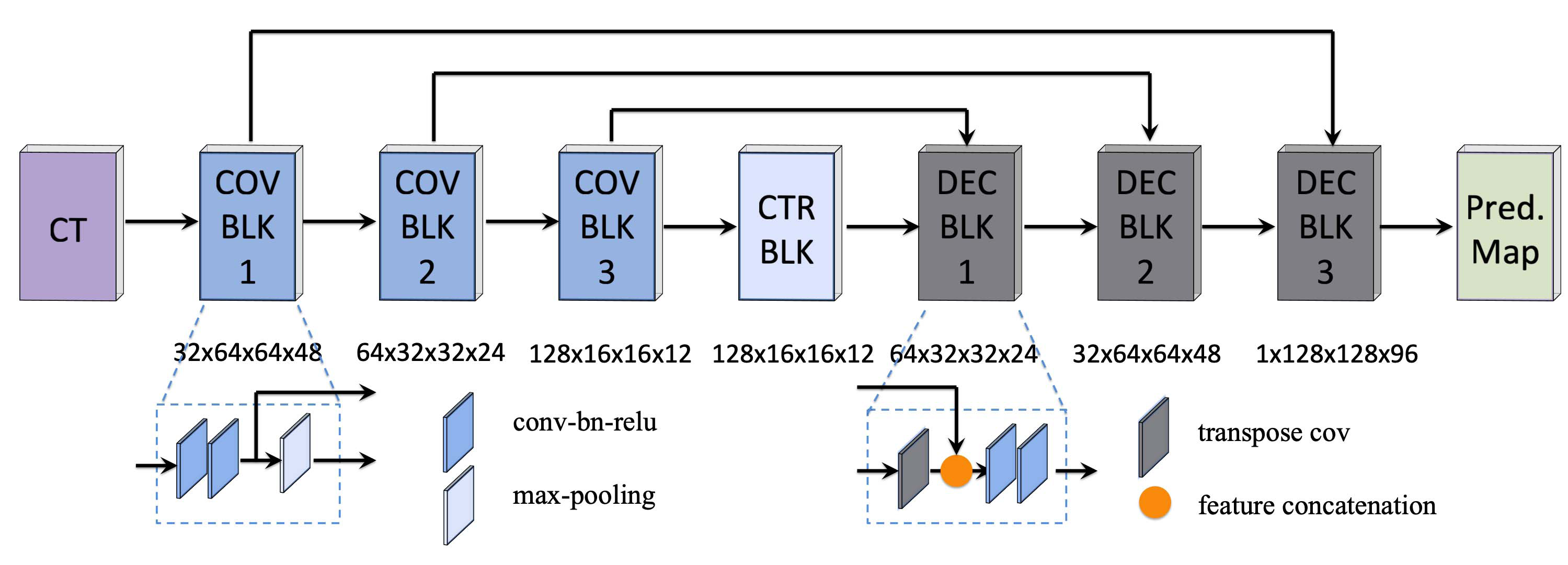} \caption{Architecture of the proposed deep regression network (DRN). DRN is
a fully 3D model that employs convolution and max-pooling to aggregate
contextual information, and uses transpose convolution and long-range
skip connection for better determination of seed locations. The numbers
under each block represent the dimensions of its output, in which
the first dimension denotes the feature channel.}
\label{fig-drn}
\end{figure}

We train a deep regression network (DRN) to map the input CT images
to the probability map using a symmetric convolutional encoding-decoding
structure, as shown in Fig. \ref{fig-drn}. Convolution and max-pooling
are employed to aggregate contextual information of CT images in the
encoding pathway, and transpose convolution is used to recover the
original resolution in the decoding pathway. Each convolutionl layer
is followed by batch normalization and rectified linear unit (ReLU)
to facilitate gradient back-propagation. Long-range skip connections,
which bridge across the encoding blocks and the decoding blocks, are
also created to allow high resolution features from encoding pathway
be used as additional inputs to the convolutional layers in the decoding
pathway. By explicitly assembling low- and high-level features, DRN
benefits from local and global contextual information to reconstruct
more precise probability map of seed locations. Considering the target
probability map is non-negative, we use $softplus$ as the activation
function in the last convolutional layer to ensure a positive output
of DRN, which approximates the ReLU function as: 
\begin{equation}
softplus(x)=\frac{1}{\beta}\cdot log(1+exp(\beta\cdot x)).
\end{equation}
In this study, we set $\beta=1$. The convolutional kernel size is
fixed as $3$ and stride as $1$, except for transpose convolution
where we set both kernel size and stride as $2$ for upscaling purpose.
Zero-padding is used to ensure the same dimension during convolution.
All the operations are performed in 3D space.

Training DRN is achieved by minimizing a loss function between the
predicted probability map $F((I\vec{x}),\,\vec{w})$ and the target
map $P(\vec{x})$. Since the majority of voxels in the target probability
map belongs to background, DRN tends to focus more on learning background
rather than the Gaussian-shaped seed annotations. In order to account
for this imbalance between background and seed annotations, we use
a weighted Mean Squared Error (MSE) as the loss function, with the
weight as the target map $P(\vec{x})$:
\begin{equation}
L(\vec{w})=\frac{1}{N}\cdot\sum_{n=1}^{N}[P(\vec{x}_{n})\cdot(P(\vec{x}_{n})-\hat{P}(\vec{x}_{n},\,\vec{w}))^{2}],
\end{equation}
where $N$ is the total number of voxels in the training mini batch.

\subsection{Implementation}

Our DRN was implemented with Python using Pytorch (v.0.4) package. Training DRN took 500 iterations from scratch
using Adam stochastic optimization method with a batch size
of $4$. The initial learning rate was set as $0.003$, and learning
rate decay and early stopping strategies were utilized when validation
loss stopped decreasing. In order to reduce overfitting, we randomly
flipped the input volume in left/right, superior/inferior, and anterior/posterior
directions on the fly for data augmentation. We used seven-fold cross
validation to evaluate the performance of our model on the training
dataset, in which a few hyper-parameters were also experimentally
determined via grid search. All the experiments were conducted on
a workstation with four Nvidia GTX 1080 TI GPUs.

As for pre-processing, we simply truncated the voxel values of all CT scans to the range of $[-80,\,175]$ HU to eliminate the irrelevant
image information. The CT images were resampled to $0.5$ mm isotropically
and $128\times128\times96$ volume of interest (VOI) centered on the
prostate was extracted from the entire CT image as input to DRN. During
inference, the new CT images were pre-processed following the same
procedure as training data preparation, then the trained DRN was applied
to VOI to yield a 3D probability map. We used a 3D watershed segmentation
algorithm to convert the probability map to the final seed locations.

\section{Experiments}

We assembled a database of $100$ prostate cancer patients treated
with seed implant brachytherapy from 2008 to 2019 in our institution.
The number of implanted seeds (Palladium 103) ranged from $48$ to
$156$. Seventy patients with $5534$ seeds were randomly selected
for model training and validation, while the remaining $30$ patients
with $2286$ seeds were reserved for independent testing. A CT scan
was performed on each patient $30$ days after implantation, with
in-plane resolution ranging from $0.6\times0.6$ to $1.4\times1.4$
mm and slice thickness from $2.5$ to $3.0$ mm.

The ground truth was obtained by a semi-automatic procedure, in which
VariSeed seed finder algorithm was first used to search implanted
seeds near prostate region in the CT images. Since this automatic
procedure usually results in a few erroneous seed placements, user
intervention was required to correct these errors based on the seed
locations in the CT images. The seed localization as well as the reconstructed
radiation dose distribution were finally approved by a radiation oncologist.

We evaluated the performance of the proposed method by comparing the
pair-wise distance between the predicted seed locations and the ground-truth
locations. For a seed obtained from the automated method and one from
ground truth, if their distance was the shortest among the list of
seeds that needed to be paired, they were considered as a pair and
removed from the list. If a pair-wise distance was smaller than $3$ mm,
the corresponding ground truth seed was considered as being correctly
identified by the automated method.

\begin{figure}[t!]
\includegraphics[width=1\textwidth]{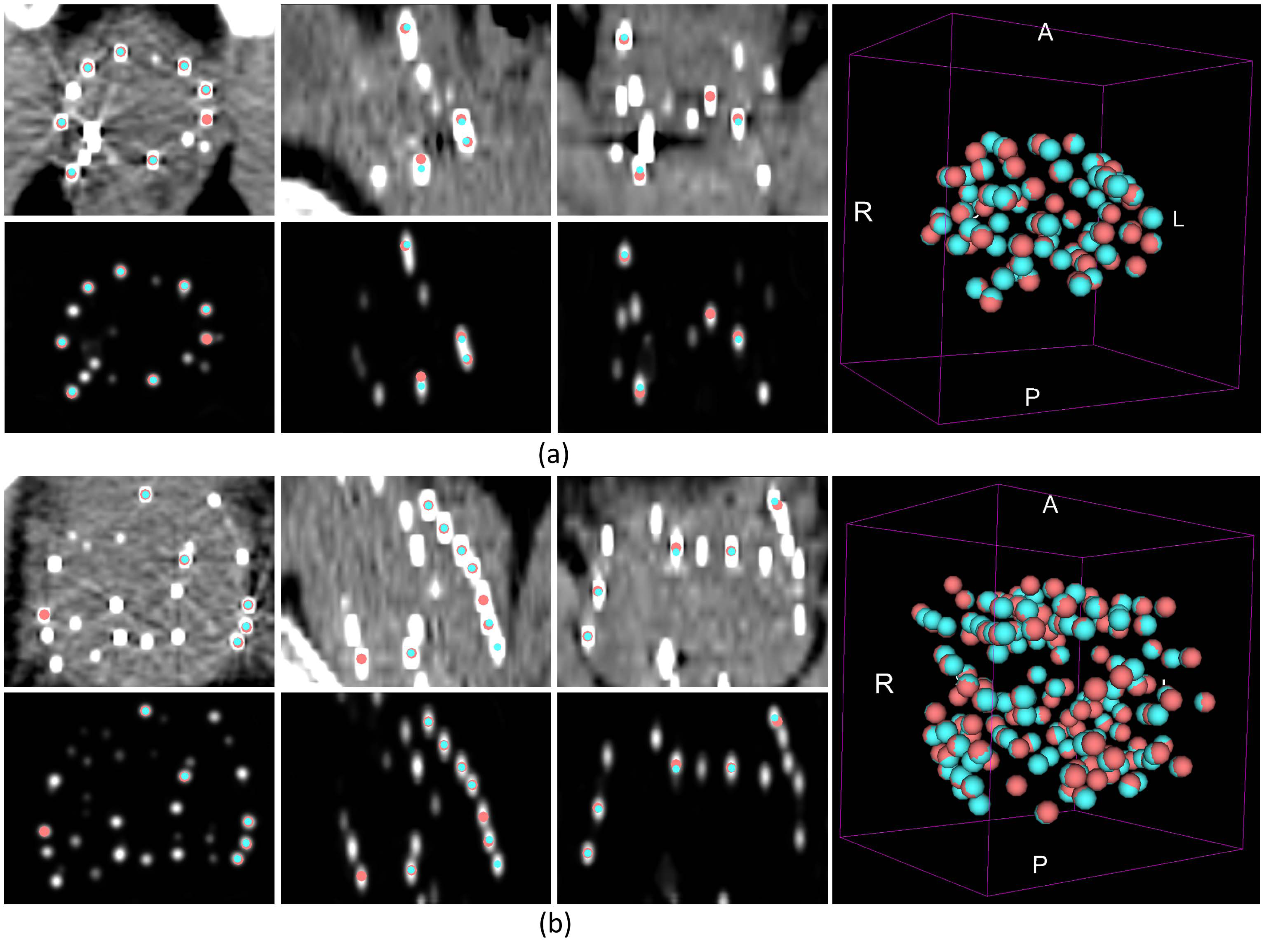} \caption{Two examples of PID study in CT images in axial, sagittal and coronal
views. The second and fourth rows are the corresponding probability
maps generated by the proposed DRN model. The right column shows the
overall 3D distributions of the ground truth and seeds identified
by DRN. In each figure, the red dots represent the ground truth while
the cyan dots are seed locations identified by DRN.}
\label{fig-results}
\end{figure}

Figure \ref{fig-results} shows two examples of PID study in CT images
in axial, sagittal and coronal views, respectively, in which $77$
seeds were implanted in patient (a) and $143$ seeds in (b). Also
shown are the corresponding DRN predictions of the probability map.
It clearly shows that the metal artifacts and seed overlap appearance
are significantly suppressed, which makes the seed localization much
easier. The plots on the right show the 3D distributions of the ground
truth and the seeds identified by DRN. Overall, it took about $60$
seconds for DRN to recover the number of implanted seeds on $30$
testing patients. The median pair-wise distance was $0.70$ mm {[}$25\%-75\%$:
$0.36-1.28$ mm{]}.

Table \ref{tab-compare} details the comparison between DRN and VariSeed
seed finder in seed detection, in which the first and fourth rows
list the number of implanted seeds. For a large range of number of
implanted seeds (from $48$ to $143$), the proposed DRN outperformed
VariSeed by a big margin on almost every patient. Overall, DRN correctly
identified 2150 out of 2286 seeds ($94.1\%$) in $30$ testing patients,
achieving $16\%$ improvement as compared to VariSeed ($81.0\%$).

\begin{table}
\caption{Comparison between DRN and VariSeed in seed detection accuracy on
the $30$ testing patients. Bold values are the numbers of implanted
seeds in each patient.}
\label{tab-compare} %
\begin{tabular}{l||c|c|c|c|c|c|c|c|c|c|c|c|c|c|c}
\hline 
\textbf{No. of seeds} & \textbf{48} & \textbf{50} & \textbf{52} & \textbf{52} & \textbf{58} & \textbf{58} & \textbf{60} & \textbf{62} & \textbf{66} & \textbf{66} & \textbf{66} & \textbf{69} & \textbf{69} & \textbf{71} & \textbf
{71}\tabularnewline
\hline 
DRN (\%) & 95.8 & 92.0 & 96.2 & 98.1 & 94.8 & 87.9 & 91.7 & 91.9 & 86.4 & 97.0 & 97.0 & 94.2 & 91.3 & 93.0 & 95.8\tabularnewline
VariSeed (\%) & 79.2 & 48.0 & 42.3 & 65.4 & 79.3 & 77.6 & 76.7 & 69.4 & 75.8 & 81.8 & 90.9 & 84.1 & 79.7 & 74.6 & 91.5\tabularnewline
\hline 
\hline 
\textbf{No. of seeds} & \textbf{72} & \textbf{72} & \textbf{74} & \textbf{77} & \textbf{78} & \textbf{79} & \textbf{82} & \textbf{84} & \textbf{88} & \textbf{95} & \textbf{99} & \textbf{100} & \textbf{108} & \textbf{117} & \textbf
{143}\tabularnewline
\hline 
DRN (\%) & 94.4 & 93.1 & 91.9 & 96.1 & 94.9 & 94.9 & 93.9 & 97.6 & 95.5 & 90.5 & 94.0 & 92.0 & 90.7 & 100.0 & 95.8\tabularnewline
VariSeed (\%) & 77.8 & 90.3 & 79.7 & 87.0 & 67.9 & 87.3 & 100.0 & 76.2 & 84.1 & 81.1 & 85.9 & 73.0 & 85.2 & 94.0 & 92.3\tabularnewline
\hline 
\end{tabular}
\end{table}

\section{Conclusion}

In this paper, we pioneered the application of deep learning in the
task of identifying radioactive seeds in CT-based post-implant dosimetry
study for patients undergoing prostate brachytherapy. Despite the
challenges in seed localization in CT images, the proposed deep regression
model achieved much higher detection accuracy as compared to a widely-used
commercial software on a large clinical database. Also, our model
was found to be very efficient, taking about 2 seconds on average
for a new test case. Instead of manually drawing 3D bounding box or
mask on each seed, our method only requires dot annotations as ground
truth for model training, which greatly simplifies the data labelling
procedure. This weakly-supervised learning framework can be easily
generalized to other object detection tasks such as fudicial marker
tracking in 2D/3D real-time imaging and source/catheter positioning
in high dose rate (HDR) brachytherapy.

\section*{Acknowledgment}

This work is partially supported by grant UL1TR001433 from the National
Center for Advancing Translational Sciences, National Institutes of
Health, USA.

%
%
%
%

\end{document}